\documentclass[12pt]{article}

\usepackage{graphics}
\usepackage{amsthm}
\usepackage{amsmath}
\usepackage{amssymb}
\usepackage{enumerate}
\usepackage{colortbl}
\usepackage{setspace}

\usepackage{latexsym}
\usepackage{amsfonts}
\usepackage{psfrag}
\usepackage{graphicx}

\usepackage{url}

\usepackage[authoryear,comma,longnamesfirst,sectionbib]{natbib}

\begin{document}
\newtheorem{prop}{Proposition}
\newtheorem{conj}{Conjecture}
%\title{\textbf{\Large Small Sample Inference for Two-way Capture-Recapture Experiments\protect\footnote{ This work was supported by a discovery grant from the Natural Sciences and Engineering Research Coucil of Canada and by the Canada Research Chair in Statistical Sampling and Data Analysis. Corresponding author: yauck.mamadou@uqam.ca }}}

\title{\textbf{\Large Small Sample Inference for Two-way Capture-Recapture Experiments}}

\author{{\normalsize \scshape{Louis-Paul Rivest} } \vspace{0.5cm}\\
\textit{\normalsize Department of Mathematics and Statistics, Universit\'e Laval}, \\
\textit{\normalsize 1045 avenue de la m\'{e}decine, Quebec, QC, G1V 0A6 Canada}\vspace{0.5cm}\\
{\normalsize \scshape{Mamadou Yauck*} } \\
\normalsize yauck.mamadou@uqam.ca \\
\textit{\normalsize Department of Mathematics, Université du Québec à Montréal}, \\
\textit{\normalsize 201 Avenue du Président-Kennedy, Montreal, QC, H3C 3P8 Canada}\vspace{0.5cm} }

\date{ }
\maketitle

\vspace{.5cm}
%{Corresponding author: Louis-Paul.Rivest@mat.ulaval.ca}

%\footnote{ This work is supported by }

\vspace{0.3cm}
\linespread{1.5}

\begin{abstract}
The properties of the generalized Waring distribution defined on the non negative integers are reviewed.  Formulas for its moments and its mode are given.  A construction as a mixture of negative binomial distributions is also presented.  Then we turn to the Petersen model for estimating the population size $N$ in a two-way capture recapture experiment. We construct a Bayesian model for $N$ by combining a Waring prior with the hypergeometric distribution for the number of units caught twice in the experiment.  Credible intervals for $N$ are obtained using quantiles of the posterior, a generalized Waring distribution.  The standard confidence interval for the population size constructed using the asymptotic variance of Petersen estimator and .5 logit transformed interval are shown to be special cases of the generalized Waring credible interval. The true coverage of this interval is shown to be bigger than or equal to its nominal converage in small populations, regardless of the capture probabilities.  In addition, its length is substantially smaller than that of the .5 logit transformed interval.  Thus the proposed generalized Waring credible interval appears to be the best way to quantify the uncertainty of the Petersen estimator for populations size.
\end{abstract}


\doublespacing

\section{Introduction}

The two-way capture recapture experiment has an interesting history that is reviewed in \citet{Lec65}.  While Petersen was the first to formulate, in 1889, a probability model for this experiment, Lecren argues that the so-called Petersen estimator for the size of the population $N$ was first proposed by Dahl, in 1917, and by Lincoln in 1930. Nowadays this model, that assumes that the results for the two capture occasions are independent, is still used in epidemiological applications, such as the evaluation of HIV prevalence \citep{gutreuter2022comparative} and the estimation of road traffic fatalities \citep{world2023road}. The determination of a confidence interval for $N$ is still open to debate.  In his Section 3.1.4, \citet{Seb82} suggests constructing such an interval using various approximations to the hypergeometric distribution for the number of units caught twice.  \citet{jensen1989confidence} shows that the distribution of Petersen estimator $\hat N$ is skewed and 100(1-$\alpha$)\% confidence intervals constructed using $\hat N \pm z_{1-\alpha/2}\sqrt{ \hat v}$, where $\hat v$ is an estimator of the asymptotic variance, have poor coverage properties; he suggests using the $1/N$ transformation.
In a large simulation study, \citet{sadinle2009transformed} compares several confidence intervals for $N$.  He concludes that the 0.5 transformed logit confidence interval, presented  in section \ref{sec:freq},  `` has an
exceptionally good performance even for extreme capture probabilities, i.e. near to 0 or 1." A recent discussion of confidence intervals for $N$ can be found in \cite{lyles2021alternative}. Bayesian methods have been studied  for experiments involving several lists, see for instance \citet{castledine1981bayesian}  and \cite{george1992capture};  their implementation in a two-way experiment does not seem to have been fully investigated.

Following  \citet{garcia2006bayesian} and \citet{webster2013estimating}, the goal of this work is to implement a proposal in \citet[Eq. 6]{seber1986review}.  He suggests creating a Bayesian model for $N$ by combining a prior for $N$ with the hypergeometric distribution for the number of units captured twice, given $N$. Credible intervals for $N$ would then be constructed using quantiles of the posterior distribution.  This paper proposes selecting the Waring distribution \citep{irwin1963place} as a prior for $N$.  This leads to a simple posterior, the generalized Waring  distribution \citep{irwin1975generalized}, whose properties have been investigated in the statistical literature, see \citet[chapter 6]{johnson2005univariate}.  The objectives of this paper are to present the prior information conveyed by the Waring distribution and to investigate the frequentist properties of credible intervals for $N$ constructed using quantiles of the generalized Waring distribution in small populations of $N=20,50$ units. The findings are that, for non-informative Waring priors, the true coverage of this interval is at least as large as its nominal coverage while its length is, in general, much smaller than that of the .5 transformed logit confidence interval. Thus the generalized Waring credible interval proposed in this work appears to be the best way to quantify the uncertainty regarding the population size in a two-way capture recapture experiment.

The outline of this paper is as follows. In Section \ref{sec:genwardist}, we present the generalized Waring distribution and its properties. New inferential procedures for Petersen's model are presented in Section \ref{Petersen}; we derived credible intervals for $N$ using quantiles from the posterior distribution for the number of missed units (Section \ref{sec:bay}) as well as asymptotic confidence intervals when $N$ goes to $\infty$ (Section \ref{sec:freq}). A simulation study comparing our credible interval to the 0.5 transformed logit confidence interval, in a frequentist setting, is presented in Section \ref{sec:sim}.

\section{The generalized Waring distribution}\label{sec:genwardist}

This section reviews properties of the generalized Waring distribution (GWD) that has been thoroughly investigated in a series of papers, see \citet{irwin1975generalized} and references therein. This distribution has a well-defined mode and a very long tail that makes it suitable for modeling accident counts \citep{xekalaki1983univariate}.  For the Petersen model, this distribution is interesting because the posterior distribution of the population size $N$ belongs to that family for the model considered in Section \ref{Petersen}. Thus, credible intervals for the unknown population size $N$ can be constructed using the quantile function of the GWD distribution.

The GWD distribution function is defined on the non-negative integers; it depends on three positive parameters, $a,b,c$ such that $c>a+b$. It is convenient to label it $GWD(a,b,c)$. The probability mass function of $Y$, a random variable with a $GWD(a,b,c)$ distribution, is given by
% Irwin's parameter a=a, k=b, \rho=c-a-b
\begin{equation}\label{waring}
\mbox{Pr}(Y=k)=p_{a,b,c}(k)= \frac 1{C_w} \frac {(a)_k(b)_k}{(c)_k}\frac 1{k!},  \quad k=0,1,\ldots,
\end{equation}
where $(a)_k=a(a+1)...(a+k-1)$, $(a)_0=1$, and $C_w$ is a normalizing constant. This probability mass function is related to the hypergeometric function $_2F_1(a,b;c;z)$ defined by
$$
_2F_1(a,b;c;z) =\sum_{k=0}^\infty \frac {(a)_k(b)_k}{(c)_k}\frac {z^k}{k!} \quad \text{for}\, |z|\le 1.
$$
For instance, the normalizing constant is $C_w={_2}F_1(a,b;c;1)=\{\Gamma(c-a-b)\Gamma(c)\}/\{\Gamma(c-b)\Gamma(c-a)\}$.  This is known as Gauss rule. Also, the probability generating function of (\ref{waring}) is $E(z^Y)={_2}F_1(a,b;c;z)/{_2}F_1(a,b;c;1)$  and the properties of the hypergeometric function can be used to derive the factorial moments of $Y$. Indeed, its mean and its variance are equal to
\begin{equation}\label{eq:moment}
E(Y)=\frac{ab}{c-a-b-1}, \quad
\mbox{Var}(Y)=\frac{ab(c-a-1)(c-b-1)}{(c-a-b-2)(c-a-b-1)^2},
\end{equation}
provided that $c>a+b+2$, see Equation (6.24) of \citet{johnson2005univariate}. \citet{irwin1975generalized} also gives the following formula for the skewness coefficient,
$$
\beta_1 =   \frac{(c+a-b-1)^2(c-a+b-1)^2(c-a-b-2)}
{ab(c-a-1)(c-b-1)(c-b-a-3)^2},
$$
provided that $c>a+b+3$, and another one for the kurtosis, see Equations (3.5) and (3.6) of \citet{irwin1975generalized} for more details. The mode of the distribution is evaluated in Equation (5.1) of \citet{irwin1975generalized}; it is equal to
\begin{equation}\label{eq:mode}
\mbox{mode}(Y)=\lfloor \frac{(a-1)(b-1)}{c-a-b+1}\rfloor.
\end{equation}
It is the largest integer that is smaller than or equal to $ (a-1)(b-1)/(c-a-b+1)$. When either $a$ or $b$ is equal to 1, the mode is at 0 and the probability function \eqref{waring} decreases with $k$.

The $GWD(1,b,c)$ is called the simple Waring distribution while $a=b=1$ gives the \citet{yule1925ii} distribution.  In this work, the Waring distribution is used as a prior distribution for the population size $N$. This distribution has interesting properties summarized in the next proposition.
\begin{prop} \label{prop1}
If $Y$ has a $GWD(1,b,c)$ distribution and $m$ is a positive integer, then
\begin{enumerate}
\item[i)] The conditional distribution of $Y-m$ given that $Y \ge m$ is a $GWD(1,b+m,c+m)$;
    \item[ii)] $\mbox{Pr}(Y \ge k)= (b)_k/(c-1)_k$.
\end{enumerate}
\end{prop}
The first result is a consequence of  (\ref{waring}) while the second can be deduced from equation (1.4) in \citet{irwin1975generalized}.

The generalized  Waring distribution can be derived as a mixture of the negative binomial distribution. The probability function of random variable $X$ distributed as a negative binomial with parameter $a>0$ and $p\in (0,1)$ is
\begin{equation}\label{eq:negbin}
\mbox{Pr}(X=k|p)=\frac{\Gamma(a+k)}{\Gamma(a)k!}p^a(1-p)^k, \quad k=0,1,\ldots.
\end{equation}
Suppose now that $p$ has a Beta distribution with parameters $c-b-a>0$ and  $b>0$, then the marginal distribution of $X$, obtained by integrating on $p$, is
\begin{eqnarray*}
\mbox{Pr}(X=k)&=&\frac{\Gamma(a+k)}{\Gamma(a)k!}\frac{\Gamma(c-b)\Gamma(b+k)\Gamma(c-a)}{\Gamma(c+k)\Gamma(c-b-a)\Gamma(b)}\\
&=& \frac{\Gamma(c-b)\Gamma(c-a)}{\Gamma(c)\Gamma(c-a-b)} \frac {(a)_k(b)_k}{(c)_k}\frac 1{k!}, \quad k=0,1,\ldots.
\end{eqnarray*}
This is equal to (\ref{waring}). Because of this construction, a possibly shifted version of the generalized Waring distribution is also known as the beta-Pascal or the beta-negative-binomial distribution in the statistical literature.  To evaluate the cumulative GWD distribution function, rather than summing the probabilities \eqref{waring}, we  calculate the expectation of a negative binomial distribution function with parameter $a>0$  and $p$ where $p$ has a beta distribution, using the integrate function in \texttt{R}.

\citet{irwin1975generalized} discusses several limiting distributions for the generalized Waring distribution.  If as $n$ goes to $\infty$ the three parameters $(a,b,c)$ are $O(n)$ in such a way that $c-b-a$ is also $O(n)$, then $\{Y-E(Y)\}/\sqrt{\mbox{Var}(Y)}$ is approximately distributed as a standard normal.  Another limiting case is obtained when $a$ is fixed and both $c-b-a$ and $b$ are  $O(n)$ as the limiting distribution is a negative binomial distribution with parameters $a$ and $p=\lim\, (c-b-a)/(c-a)$.

\section{New inference methods for Petersen's model} \label{Petersen}

This section reviews the sampling model underlying the Petersen estimator.  A population contains an unknown number $N$ of units and each can be detected by two sources.
The data set contains frequencies $n_{10}$ and $n_{01}$, the number of units detected by respectively the first and second source only, and $n_{11}$ the units found by the two sources. A total of $n_{\bullet \bullet}=n_{10}+n_{01}+n_{11}$ units is detected. The unknown number of units missed in the experiment can be expressed as $K=N-n_{\bullet \bullet}$. The model for $\{n_{11},n_{10},n_{01}, K\}$ underlying the Petersen estimator is a multinomial with parameter $N$ and probabilities $\{p_1p_2,p_1(1-p_2),(1-p_1)p_2,(1-p_1)(1-p_2)\}$, where $p_1,p_2\in (0,1)$ are the detection probabilities for the two sources.\\

The sufficient statistics for the nuisance parameters $p_1$ and $p_2$ are $n_{1\bullet}=n_{10}+n_{11}$ and $n_{\bullet 1}=n_{01}+n_{11}$.  Estimators for $N$ can be constructed using the conditional distribution of $n_{11}$ given $(n_{1\bullet}, n_{\bullet 1})$. This is the hypergeometric distribution whose probability function is given by
\begin{eqnarray}\nonumber
\mbox{Pr}(n_{11}|N, n_{1\bullet}, n_{\bullet 1})&=& \frac{\displaystyle {n_{\bullet 1} \choose n_{11}}{N-n_{\bullet 1} \choose n_{1\bullet}- n_{11}}}{\displaystyle {N\choose n_{1\bullet}}}\\
&=&  \frac{n_{1 \bullet }!n_{ \bullet 1}! }{n_{10}!n_{01}!n_{11}!} \label{hyper}
 \frac{(N-n_{\bullet 1})!(N-n_{1 \bullet })!}{N!(N-n_{\bullet \bullet})!}
\end{eqnarray}
for $n_{11}=\min(0, N-n_{\bullet 1}- n_{1\bullet}),\ldots, \max(n_{\bullet 1},n_{1\bullet})$, see \citet{Seb82}. One can reexpress \eqref{hyper} in terms of $K=N-n_{\bullet \bullet}$ and of the $GWD$ probability function, $p_{a,b,c}(k)$ given in \eqref{waring}, as
\begin{equation}\label{eq:distK}
\mbox{Pr}(n_{11}|N, n_{1\bullet}, n_{\bullet 1})= p_{n_{01}+1,n_{10}+1,n_{\bullet\bullet}+1}(K)
\frac{n_{\bullet 1}n_{1 \bullet}}{n_{11}(n_{11}-1)}.
\end{equation}
Since our goal is to estimate $K=N-n_{\bullet \bullet}$, \citet{seber2019brief} proposed to look at Equation (\ref{eq:distK}) as a distribution for the unknown $K$ given the data $\{n_{1\bullet}, n_{\bullet1}, n_{11}\}$.  This gives the $GWD(n_{01}+1,n_{01}+1,n_{\bullet \bullet} +1)$, a result first noted by \citet{garcia2006bayesian} and \citet{webster2013estimating}.
\\
\subsection{Some Bayesian estimation procedures  } \label{sec:bay}

Seen from a Bayesian point of view, the $GWD(n_{01}+1,n_{01}+1,n_{\bullet \bullet} +1)$ distribution in \eqref{hyper} is the posterior for $K$ that corresponds to a degenerate prior where all the positive integers receive the same probability mass.  To broaden the class of estimators for $N$, we suggest to construct priors for $K$ with a Yule distribution with parameter $\ell\ge 0$, see \citet{fienberg1999classical} for a related proposal.  This distribution can be expressed as a $GWD(1,1,1+\ell)$. Given that $N\ge n_{\bullet \bullet}$  the  prior for $K=N-n_{\bullet \bullet}$ is a Waring distribution, given by $GWD(1,n_{\bullet \bullet}+1,n_{\bullet \bullet}+1+\ell)$ using Proposition \ref{prop1}, i). Using \eqref{eq:moment}, its prior expectation is equal to $(n_{\bullet\bullet}+1)/(\ell-2)$ provided that $\ell>2$. The value $\ell=1$ gives an improper prior with a  probability for $K=k$ proportional to $1/(n_{\bullet \bullet}+ k+1)$  while for $\ell=2$  the prior's expectation is infinite. Non integer values of $\ell$ are possible; for instance $\ell=2.2$ gives a prior expectation for the number of units missed equal to the five times the number caught while the prior variance is equal to $\infty$.
To help convey the prior information associated to the Waring $GWD(1,n_{\bullet \bullet}+1,n_{\bullet \bullet}+1+\ell)$ distribution, Table \ref{tab:pri} gives the prior 95th percentiles corresponding to various values of $\ell$ and of $n_{\bullet \bullet}$.
It shows that the $\ell=2$ prior distribution is almost non-informative when compared to that for $\ell=3$.

\begin{table}[ht]
\centering
\caption{Prior 95th percentile of $K=N-n_{\bullet \bullet}$ for $\ell=2,\ 2.2,\ 3$ and various values of $n_{\bullet \bullet}$ }
\label{tab:pri}
\begin{tabular}{lllllll}
$n_{\bullet \bullet}$ &  1 & 2 & 5  &  10 & 20 & 50 \\
$\ell=2$ &  38 & 57  & 114 &  208 & 398 &  969 \\
$\ell=2.2$ & 23 & 34  & 67 &  128 & 235 &  569 \\
$\ell=3$ &   8 & 11  & 22  &  39  &  74 &  178 \\
 \hline
\end{tabular}
\end{table}

The posterior for $K$ corresponding to a $GWD(1,n_{\bullet \bullet}+1,n_{\bullet \bullet}+1+\ell)$ prior is the product of \eqref{hyper} times the prior probability function:
\begin{eqnarray}\nonumber
\mbox{Pr}(K=k|n_{11},n_{1\bullet},n_{\bullet 1})&\sim& \frac{(n_{01}+1)_k(n_{10}+1)_k}{(n_{\bullet \bullet}+1)_kk!} \frac{(n_{\bullet \bullet}+1)_k}{(n_{\bullet \bullet}+1+\ell)_k}\\
&=&\frac{(n_{01}+1)_k(n_{10}+1)_k}{(n_{\bullet \bullet}+1+\ell)_kk!} ,\ k=0,1,\ldots
\label{posterior}
% \frac{ (n_{10}+k)!(n_{01}+k)!}{(n_{\bullet \bullet}+\ell+k)!k!},
\end{eqnarray}
In other words, the posterior distribution for $K=N-n_{\bullet \bullet}$ is a $GWD(n_{01}+1,n_{01}+1,n_{\bullet \bullet}+\ell +1)$ and the Waring distribution is the conjugate prior for parameter $N$. \citet{roberts1967informative} and
 \citet{freeman1973seq} also obtained a generalized Waring posterior for $N$ in a sequential setting where the prior for $N$ depends on $n_{1\bullet}$, the number of units captured by the first source.

Using \eqref{eq:moment} the posterior expectation of $K$, the number of units missed is $(n_{01}+1)(n_{10}+1)/(n_{11}-2+\ell)$; this agrees with Equation (19) of \citet{webster2013estimating} when $\ell=0$. The mode of this distribution is $\mbox{mode}(Y)=\lfloor n_{10}n_{01}/(n_{11}+\ell)\rfloor$, see \eqref{eq:mode}. This gives the standard Petersen estimator when $\ell=0$ and Chapman bias corrected estimator \citep{Cha51} when $\ell=1$. In a similar way, the posterior variance of $K$ is found to be equal to
$$
\mbox{Var}(K) = \frac{(n_{01}+1)(n_{10}+1)(n_{1\bullet}-1+\ell)(n_{\bullet1}-1+\ell) }{(n_{11}+\ell-2)^2(n_{11}+\ell-3)}, \ n_{11}+\ell> 3.
$$
This is, when $\ell=2$, larger than the Seber-Wittes bias corrected variance estimator for the Petersen estimator \citep[chap 3.3]{Seb82}.

For a given value of $\ell$, a $100(1-\alpha)\%$ credible interval for $N$ is calculated using the quantile function, $q_{\alpha, a,b,c}$ of the generalized Waring distribution defined as the smallest integer $m$ for which Pr$\{GWD(a,b,c)\le m\}\ge \alpha $. The $100(1-\alpha)\%$ credible interval for $N$ is given by:
\begin{eqnarray}
\nonumber
\lefteqn{\mathcal{C}_{100(1-\alpha),\ell}(N)}\\
&=& \{n_{\bullet \bullet}+q_{\alpha/2,n_{01}+1,n_{01}+1,n_{\bullet \bullet}+\ell +1},
n_{\bullet \bullet}+q_{1-\alpha/2,n_{01}+1,n_{01}+1,n_{\bullet \bullet}+\ell +1}\}.
\label{eq:wci}
\end{eqnarray}
%This interval can be interpreted either as a confidence interval for $N$ in a frequentist setting or as a Bayesian credibility interval constructed with a Yule prior with parameter $\ell$.
When $\ell\ge 2$, $\mathcal{C}_{100(1-\alpha),\ell}(N)$ is defined even when $n_{11}=0$. When there are no recaptures, the posterior distribution for $K$ takes large values and provides a very wide interval for $N$. In the sequel, we do not consider the $\ell=0$ and the $\ell=1$ estimators as they lead to credible intervals for $N$ that are much wider than the $\ell=2$ intervals. The $\ell=2$ Waring prior is, for all practical purposes, non-informative;  we also present results obtained with the $\ell=3$ prior that assumes that about 50\% of the units have been caught and has an
infinite variance. 

\subsection{Approximate confidence intervals for $N$} \label{sec:freq}

When the population size $N$ goes to $\infty$, the frequencies  $\{n_{11},n_{10},n_{01}\}$ go to $\infty$ under Petersen multinomial model.  The generalized Waring quantiles can then be approximated by normal quantiles:
\begin{eqnarray*} \lefteqn{
q_{\alpha,n_{01}+1,n_{01}+1,n_{\bullet \bullet}+\ell+1}}\\
&=&
\frac{(n_{01}+1)(n_{10}+1)}{n_{11}+\ell} + z_\alpha \sqrt{\frac{(n_{01}+1)(n_{10}+1)(n_{\bullet 1}-1+\ell)(n_{1\bullet}-1+\ell)}{(n_{11}+\ell-2)^2(n_{11}+\ell-3)}}.
\end{eqnarray*}
Neglecting terms that are $O_p(1)$, the confidence interval constructed with the geneneralized Waring distribution is approximately equal to
\begin{equation}\label{eq:petci}
n_{\bullet \bullet} + \frac{n_{01}n_{10}}{n_{11}}\pm z_{1-\alpha/2}
\sqrt{\frac{n_{01}n_{10}n_{\bullet 1}n_{1\bullet}}{n_{11}^3}}
\end{equation}
in large samples.  This is the standard Wald confidence interval for $N$, constructed using the asymptotic distribution for the Petersen estimator, see \citet{sadinle2009transformed}.  Thus, the proposed Bayesian approach yields credible intervals with good frequentist coverage properties in large samples since they are similar to the standard Wald confidence interval.

Considering the skewness of the GWD distribution, its convergence to the normal distribution might be slow. The normal distribution might provide a better approximation to the distribution function of $\log K$.  The asymptotic variance of this variable, neglecting $O_p(1)$ terms, is
\begin{eqnarray*}
\sigma_K^2&=&\left(\frac{n_{11}}{n_{01}n_{10}}\right)^2 \times \frac{n_{01}n_{10}(n_{0 1}+n_{11})(n_{10}+n_{11})}{n_{11}^3} \\
  &=&\frac{(n_{01}+n_{11})(n_{10}+n_{11})}{n_{11}n_{01}n_{10}} \\
  &=& \frac 1{n_{11}} + \frac 1{n_{10}} + \frac 1{n_{01}} + \frac {n_{11}}{n_{10}n_{01}},
\end{eqnarray*}
while its asymptotic mean is $\log n_{10} + \log n_{01} -\log n_{11}$.  This leads to the following asymptotic confidence interval for $N$:
$$
n_{\bullet \bullet}+ \frac{n_{01}n_{10}}{n_{11}}\exp\left(\pm z_{\alpha/2} \sqrt{\sigma_K^2}\right).
$$
This is very close to \citet{sadinle2009transformed} $100(1-\alpha)\%$ 0.5 transformed logit (Tlogit) confidence interval that is given by
\begin{equation} \label{eq:tlogit}
n_{\bullet \bullet}+ \frac{(n_{01}+0.5)(n_{10}+0.5)}{n_{11}+.5}\exp\left(\pm z_{\alpha/2} \sqrt{\sigma_{*K}^2}\right)-0.5,
\end{equation}
where $\sigma_{*K}^2$  is calculated by adding 0.5 to the three frequencies $(n_{10},n_{01},n_{11})$ in the formula for $\sigma^2_K$.  This Tlogit confidence interval is important because it is the only confidence interval, among all those tested in \citet{sadinle2009transformed}, whose true coverage was larger than the nominal 95\% target for all test cases.

\subsection{Some examples} \label{sec:exa}
The goal of the section is to compare the GWD credible interval given in Equation \eqref{eq:wci} to the Tlogit interval given in Equation \eqref{eq:tlogit}.  In Table \ref{tab:Seb82} the first two rows are about the size of underground ant (\textit{Lasius flavus}) colonies while the third and fourth ones are concerned with daily redpolls (\textit{Acanthis linaria}) abundance; the capture recapture data for the four experiments are taken from Tables 3.7 and 3.11 in \citet[chap 3.3]{Seb82} who argues that they can analyzed using Petersen model.
%with the goal of providing confidence intervals for the number of ant workers across the six colonies. The mark-recapture experiment was conducted as follows. On both sampling occasions, colonies of ants were found by first placing a flat stone on top of the mound then wait for a period of time before lifting it up, removing and tagging the ants before returning them to the mound. 
Table \ref{tab:Seb82} gives credible intervals for $N$ calculated using the Waring prior distributions with parameters $\ell=2,3$ and the Tlogit confidence interval given by \eqref{eq:tlogit}.  An estimate of the asymmetry coefficient $\beta_1$ of the underlying generalized Waring distribution, when $\ell=0$, is also provided to evaluate the skewness of the underlying GWD distribution.

\begin{table}[ht]
\centering
\caption{Some 95\% credible and confidence intervals (Lb, Ub) for $N$ obtained with Waring's prior (with corresponding skewness parameter $\beta_1$) and the 0.5 transformed logit method.}
\label{tab:Seb82}
\begin{tabular}{rrrrrrrrrr}
 \multicolumn{3}{c}{Data}   &   \multicolumn{2}{c}{$\ell=2$} &  \multicolumn{2}{c}{$0.5$T-logit}   &   \multicolumn{2}{c}{$\ell=3$} & $\beta_1$\\
 $n_{10}$&  $n_{01}$&  $n_{11}$& Lb & Ub &  Lb & Ub & Lb & Ub& \\ % .5 logit
   \hline
493 & 142 & 7  &    5378  & 21421  & 5116 &20291 & 4968  & 18107  & 7.1 \\ % 5116 20291
%483 & 172& 17 &     3628 & 8715   & 1.3         &  3628 & 8715                \\
511 & 232 & 89 &     1853 & 2565   & 1852 & 2562 &  1843 & 2545   & 0.2 \\ % 1852 2562
  1 &   7 &  5 &     13   & 26     & 13   & 35   &  13   & 24     & 40.2 \\  % 13 35
  10 &  3 &  3 &     17   & 69     & 17   & 74   &  17   & 54     & NA  \\ % 17 74
   \hline
\end{tabular}
\end{table}

%   \hline
%493 & 142 & 7  & 6517 & 33128 & 5882 & 26102 & 5379  & 21421  & 7.1 \\ % 5116 20291
%%483 & 172& 17 & 3930 & 10037 & 3771 & 9329  &  3628 & 8715   & 1.3 \\
%511 & 232 & 89 & 1875 & 2608 & 1864  & 2586  &  1853 & 2565   & 0.2 \\ % 1852 2562
%  1 &   7 &  5 & 13   & 36   & 13    & 30    &  13   & 26     & 40.2 \\  % 13 35
%  10 &  3 &  3 & 19  & 117   & 18    & 103   &17     & 69     & NA  \\ % 17 74
%   \hline

The first data set treated in Table \ref{tab:Seb82} has few recaptures.  With $n_{\bullet \bullet}=642$,  the 97.5th percentile of the $GWD(1,643, 645)$ prior is 25 076.  This is close to the upper bound of the $\ell=2$ 95\% confidence interval;  this highlights that the 7 recaptures in this data set are not really informative.  The 95\% credible interval for $\ell=2$  in Table \ref{tab:Seb82}, (5379,\ 21421) is much wider than the standard 95\% confidence interval constructed using \eqref{eq:petci} which is given by (3470,\ 15316).  This was expected considering the large skewness ($\beta_1=7.1$)  of
the underlying GWD distribution.  For the second data set of Table \ref{tab:Seb82},  the skewness is close to 0 ($\beta_1=0.2$) and the three 95\% credible/confidence intervals are close to (1803,\ 2495), that obtained using \eqref{eq:petci}. In Table \ref{tab:Seb82}, the Tlogit confidence intervals are similar to the $\ell=2$  GWD interval, except for the two data sets with small frequencies where they are much larger.  This suggests that the log-normal approximation to the large GWD percentiles overestimates their true values in small samples.

\begin{figure}
\centering
\includegraphics[scale=0.7]{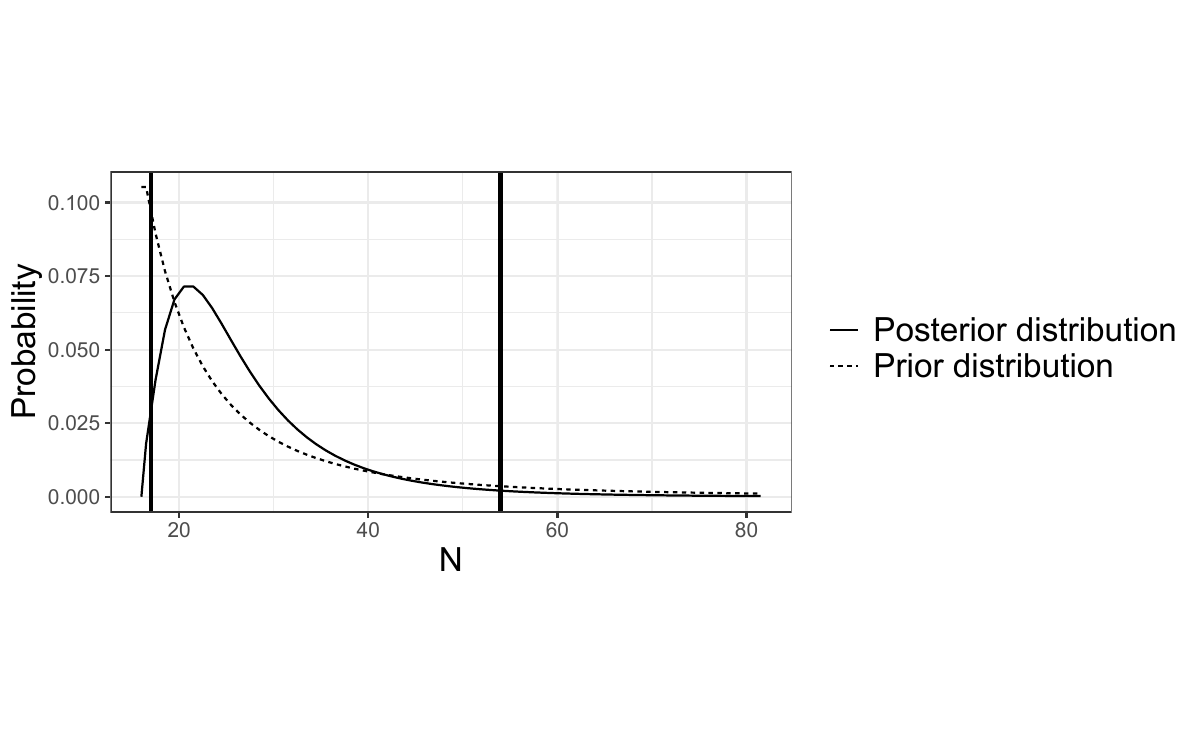}
\vspace{-2.8cm}
\caption{Prior and posterior distributions for $N$ corresponding to the fourth example in Table \ref{tab:Seb82}. Lower and upper bounds for the 95\% credible interval are represented by vertical lines.}
\label{fig:fourthexample}
\end{figure}

Figure \ref{fig:fourthexample} considers the fourth example in Table \ref{tab:Seb82}.  For this data set, the $\ell=2$ posterior distribution for $N$ is the GWD(11,4,19) distribution. Considering \eqref{eq:mode}, the posterior mode is $\hat N=21$; the prior and the posterior distribution for $N$ are shown in Figure \ref{fig:fourthexample}.  This figure is similar to one produced by an R shiny app implementing the GWD credible intervals for Petersen model proposed in this work, see \cite{RivYauckShiny23}.

\section{The expected length and the coverage of the GWD credible interval for $N$}\label{sec:sim}

\begin{figure}
\centering
\includegraphics[scale=0.7]{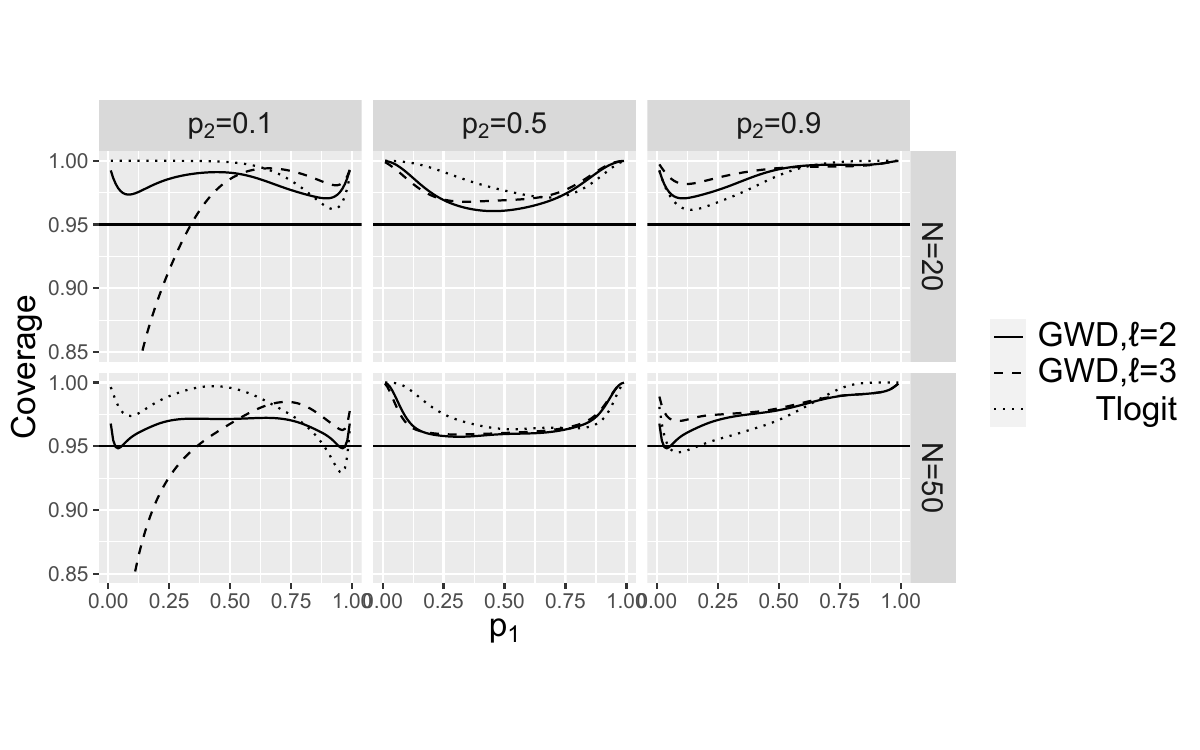}
\vspace{-2cm}
\caption{Expected coverage, as a function of $p_1, p_2$, of three 95 \% credible/confidence intervals for $N=20$ and $N=50$:  the Tlogit interval and the GWD intervals for $\ell=2,3$.}
\label{fig:cov}
\end{figure}

This section evaluates the performance of the GWD credible interval using frequentist methods; the population size $N$ is set to 20 and 50.  For a population of size $N$, there are ${ (N+3) \choose 3}$ possible capture-recapture samples. For each one, a 95\% credible/confidence interval for $N$ can be calculated using one of the methods discussed in Section \ref{sec:exa}.  The expected coverage of an interval is the expectation of a dichotomous variable that takes the value 1 if it contains the populations size $N$ and 0 otherwise. Its value depends on the detection probabilities $p_1$ and $p_2$.  Figure \ref{fig:cov}  shows the coverage of several credible/confidence intervals as a function of $p_1$ and $p_2$, for $N=20,50$.
%When $\ell=1$, the GWD confidence interval ia not well defined if $n_{11}=0$ because the constraints $c>b-a$ for the GBD parameters fails. In these situations the confidence interval length is set to $\infty$;  it therefore covers the true value $N=20$.

Figure \ref{fig:cov} reveals that the true coverage of the GWD credible interval is larger than the nominal value of 95\% for $\ell=2$ for almost all the pairs $(p_1,p_2)$ considered. It performs as well as the Tlogit confidence interval.  For $\ell=3$ the GWD coverage when $p_2=0.1$ and $p_1<.3$ is well below 95\%.  This could be expected considering the discussion in Section \ref{sec:bay}.  In these cases the GWD prior expectation of $K$, $n_{\bullet \bullet}+1$, is in general much smaller than the true number of units that have not been captured. The Waring $\ell=3$ prior underestimates the uncertainty for $N$.

\begin{figure}
\centering
\includegraphics[scale=0.57]{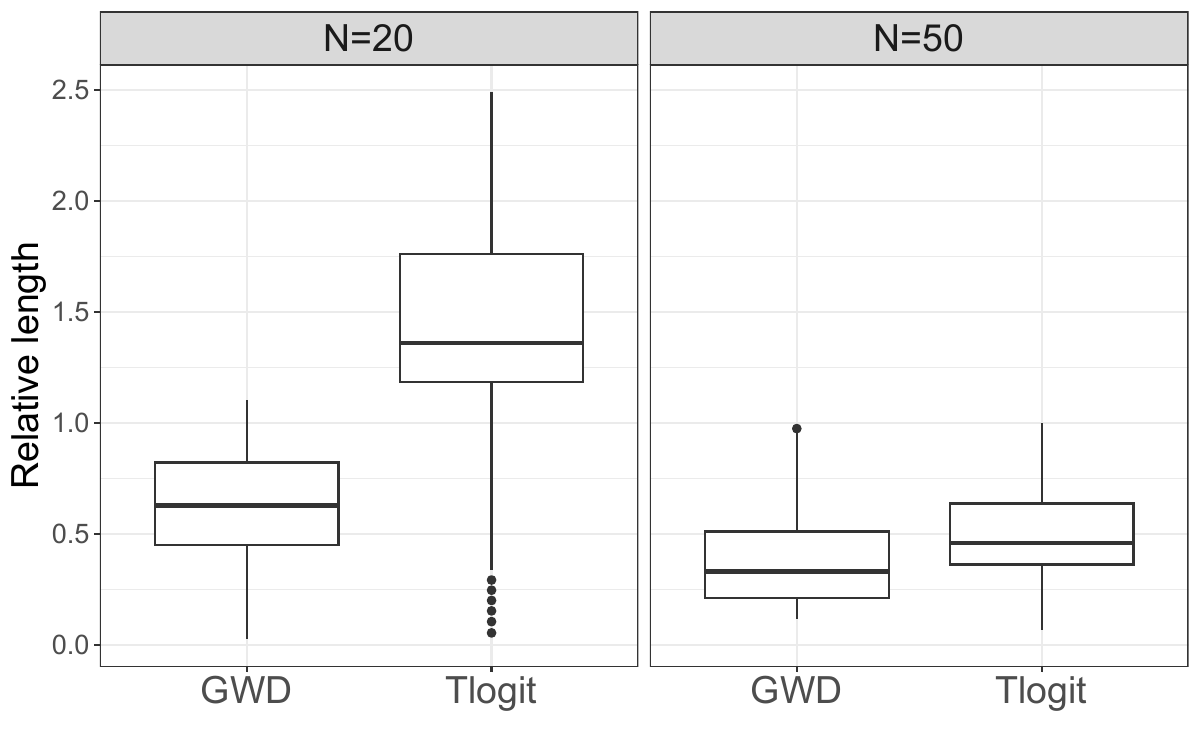}
\vspace{-0.7cm}
\caption{Boxplots for the relative lengths of Tlogit confidence intervals and the corresponding GWD credible intervals, with $\ell=2$, at the 95\% confidence level. This representation only visualizes the distribution of the relative lengths when the interval length is less than 50 for both methods.}
\label{fig:len}
\end{figure}

The next step is to compare the relative lengths of the GWD, with $\ell=2$, and the Tlogit credible/confidence intervals. First samples with $n_{11}$ equal to 0 or 1 are excluded, as not containing much information about population size.  For the remaining ${ (N+1) \choose 3}$  samples, the relative lengths, e.g. the lengths divided by the population size, of the two 95\% intervals under study were evaluated.  Figure \ref{fig:len} shows boxplots of the $\ell=2$ GWD and the Tlogit relative lengths, for $N=20,50$.  The samples for which the length of at least one of the two intervals  is larger than 50  have been discarded.

In Figure \ref{fig:len}, the length of the Tlogit 95\% confidence is larger than the corresponding $\ell=2$ GWD interval  for 94.4\% of the samples when $N=20$.  Indeed the GWD credible interval lengths are less than half that of the Tlogit interval when $N=20$.  When $N=50$ the GWD intervals retain their superiority  even if the differences are smaller. For larger values of $N$, the $\ell=2$ GWD and the Tlogit intervals give similar results, as illustrated in the second examples of Table \ref{tab:Seb82}.

\section{Discussion}

Generalizations of the methodology presented in this work to experiments with $I>2$ capture occasions is not straightforward. A generalization of Petersen's model is the class of log-linear models proposed in \citet{Fie72}.  These models have sufficient statistics for the nuisance parameters and the conditional distribution of the data given those sufficient statistics is sometimes known. For instance, when there are $I=3$ capture occasions, consider the model with an interaction between occasions 2 and 3. The conditional distribution of the data, given the sufficient statistics, is proportional to \eqref{hyper} and the GWD credible interval applies to this experiment. It suffices to combine capture occasions 2 and 3 in a single capture occasion.   In the model of independence between $I>2$ capture occasions, the conditional distribution, given the sufficient statistics for the $I$ capture probabilities, is an extended hypergeometric distribution discussed in \citet{Dar58}.  \citet{garcia2006bayesian} looked at this distribution as one for the population size $N$ given the data.    This posterior involves a fairly complex  ${_I}F_{I-1}$ hypergeometric function that does not seem to have been investigated in the statistical literature. Thus, except in some special cases, the methodology presented in this paper does not apply to experiments involving more than two capture occasions and the evaluation of a Bayesian posterior distribution for $N$ needs to be done through a simulation algorithm.

The calculation of most credible intervals for $N$ in Petersen's model relies on quantiles of the normal distribution, see \citet{sadinle2009transformed} and \citet{jensen1989confidence}. GWD credible intervals are more complex as they involve the GWD distribution.  To ease the evaluation of these confidence intervals, R-codes for evaluating the probability mass function \eqref{waring}, the cumulative distribution function and the quantile function of the GWD distribution are provided in the supplementary material.  An R-shiny implementing these calculations, \textit{ShinywaringCapt}, is also available \citep{RivYauckShiny23}.

\section*{Acknowledgments}
This work was supported by a discovery grant from the Natural Sciences and Engineering Research Coucil of Canada and by the Canada Research Chair in Statistical Sampling and Data Analysis.

\bibliographystyle{jasaauthyear}
\bibliography{RivestSSC}

\end{document}